\documentstyle[epsfig,12pt,aps]{revtex}
\begin{document}
\newcommand{\p}{\partial}
\newcommand{\ls}{\left(}
\newcommand{\rs}{\right)}
\newcommand{\beq}{\begin{equation}}
\newcommand{\eeq}{\end{equation}}
\newcommand{\beqa}{\begin{eqnarray}}
\newcommand{\eeqa}{\end{eqnarray}}
\newcommand{\bdm}{\begin{displaymath}}
\newcommand{\edm}{\end{displaymath}}
\newcommand{\fps}{f_{\pi}^2 }
\newcommand{\mks}{m_{{\mathrm K}}^2 }
\newcommand{\ms}{m_{{\mathrm K}}^{*} }
\newcommand{\mk}{m_{{\mathrm K}} }
\newcommand{\msq}{m_{{\mathrm K}}^{*2} }
\newcommand{\rhos}{\rho_{\mathrm s} }
\newcommand{\rhob}{\rho_{\mathrm B} }
\title{The nuclear equation of state probed by 
$K^+$ production in heavy ion collisions
\footnote{Proceedings to the XXXIX Interantional Winter Meeting 
on Nuclear Physics, Bormio, Italy, 2001}}
\author{C. Fuchs, Amand Faessler, S. El-Basaouny, E. Zabrodin}
\address{
Institut f\"ur Theoretische Physik der Universit\"at T\"ubingen, 
Auf der Morgenstelle 14, D-72076 T\"ubingen, Germany
}
\maketitle  
\begin{abstract}
The dependence of  $K^+$ production on the nuclear equation of state 
is investigated in heavy ion collisions. 
An increase of the excitation function of $K^+$ 
multiplicities obtained in heavy ($Au+Au$) over light ($C+C$) systems 
when going far below threshold which has been observed by the 
KaoS Collaboration strongly favours a soft equation of state. This 
observation holds despite of the influence of an in-medium kaon 
potential predicted by effective chiral models which is necessary 
to reproduce the experimental $K^+$ yields. Phase space effects 
are discussed with respect to the $K^+$ excitation function.
\end{abstract}
\section{Introduction}
From the very beginning kaons have been considered as one of the 
best probes to study dense and hot nuclear matter formed 
in relativistic heavy ion collisions \cite{AiKo85}. In particular at 
incident energies below the corresponding 
production thresholds in free space  $K^+$ mesons are created
in the early and
high density phase of such reactions and -- due to strangeness 
conservation -- are not reabsorbed by the nuclear environment. 
Furthermore, there exist strong evidences that kaons change their 
properties inside the nuclear medium as predicted by effective 
chiral models \cite{kapla86,weise93}. The investigation of a 
partial restoration of chiral symmetry in dense matter probed 
by $K$ mesons has strongly stimulated both experimental and 
theoretical efforts in the recent years 
\cite{li95,kaos94,fang94,fopi95,kaos99,li97,wang97,cassing99}. 

The original motivation to study the kaon production 
in heavy ion reactions at intermediate energies, namely to extract 
information on the nuclear equation of state (EOS) at high densities is 
a matter of current debate. Already in the first theoretical 
investigations by transport models it was noticed that the $K^+$ 
yield reacts sensitive on the nuclear equation of state 
\cite{AiKo85,qmd93,li95b,baoli94}, i.e. it 
was found to be about a factor 2--3 larger when a 
soft EOS was applied compared 
to a hard EOS. At that time the available data \cite{kaos94} 
already favoured a soft equation of state. However, calculations 
as well as the experimental data were still burdened with 
large uncertainties. It was further noticed \cite{qmd93} that 
the influence of the repulsive momentum dependent 
part of the nuclear interaction leads to 
a strong suppression of the kaon abundances. 
An underprediction of $K^+$ yields using more realistic 
momentum dependent forces was due to the fact that 
the production mechanism is twofold:  
baryon induced processes $ BB\longrightarrow BYK^+ $ 
where the kaon is created via binary baryon--baryon collisions 
($B$ stands either for a nucleon or a $\Delta$--resonance and $Y$ for 
a $\Lambda$ or a $\Sigma$ hyperon, respectively) and processes 
$\pi B\longrightarrow YK^+ $ induced by pion absorption. 
In the early studies on subthreshold $K^+$ production 
only the baryon induced channels have been considered 
\cite{qmd93,li95b}. As shown in 
\cite{baoli94,fuchs97,brat97} 
the pionic channel plays an important role, 
particular in heavy systems. Taking this fact into account 
the kaon yield could be explained adopting 
realistic momentum dependent nuclear interactions 
\cite{li97,cassing99,fuchs97,brat97}. 
However, the dependence of the kaon production on the nuclear EOS 
turned now out to be too small for definite conclusions.

The a recent work \cite{fuchs01} we studied the question if in the meantime 
decisive information on the nuclear EOS can be extracted from subthreshold 
kaon production in heavy ion collisions. There are several reasons 
why it appears worthwhile to do this: Firstly, there has been 
significant progress in the recent years towards a more precise 
determination of the elementary kaon production cross sections 
\cite{sibirtsev95,tsushima99}, based also on new data points form 
the COSY-11 for the reactions 
$pp \longrightarrow p K^+ X$ very close to threshold \cite{cosy11}. 
Secondly, the KaoS Collaboration has performed systematic 
measurements of the $ K^+$ production far below threshold in 
heavy ($Au+Au$) and light ($C+C$) systems \cite{sturm00}. Looking at the 
ratios built from heavy and light systems possible uncertainties which 
might still exist in the theoretical calculations should cancel out 
to a large extent which allows to draw reliable conclusions. 
Furthermore, far below threshold the kaon production is a highly 
collective process and a particular sensitivity to the compression of the 
participant matter is expected. 
\section{The Model}
The present investigations are based on the 
Quantum Molecular Dynamics (QMD) transport model \cite{ai91}. 
For the nuclear EOS we adopt soft and hard Skyrme forces 
corresponding to a compression modulus of 
K=200 MeV and 380 MeV, respectively, and with a momentum dependence 
adjusted to the empirical optical nucleon-nucleus potential 
\cite{ai91}. The saturation 
point of nuclear matter is thereby fixed at $E_B = -16$ MeV and 
$\rho_{\rm sat}= 0.17~{\rm fm}^{-3}$ \cite{ai91}. 
The calculations include $\Delta$ and $N^*(1440)$ resonances with \cite{Hu94}. 
The QMD approach with Skyrme interactions is well tested, 
contains a controlled momentum dependence and provides 
a reliable description of the reaction dynamics in the SIS 
energy range, expressed e.g. by collective nucleon flow observables as well 
as particle production. 
Also the EOS predicted by microscopic approaches (G-matrix) 
\cite{dbhf} is similar to the soft version of the 
Skyrme interaction for densities up to $2\rho_{\rm sat}$.

We further consider the influence of an 
in-medium kaon potential based on 
effective chiral models \cite{kapla86,weise93,li95,li97,brown962}. 
The $K^+$ mean field consists of a repulsive vector part 
$V_\mu = 3/8 f^{*2}_{\pi} j_\mu$ and an attractive scalar part 
$\Sigma_S = m_{{\rm K}} - \ms = m_{{\rm K}} - 
\sqrt{ \mks - \frac{\Sigma_{\mathrm{KN}}}{\fps} \rhos 
     + V_\mu V^\mu }$. Here $j_\mu$ is the baryon vector current 
and $\rhos$ the scalar baryon density and  
$\Sigma_{\mathrm{KN}} = 450$ MeV. Following \cite{brown962} in 
the vector field the pion decay constant in the medium 
$f^{*2}_\pi = 0.6 \fps$ is used. However, 
the enhancement of the scalar part using $f^{*2}_\pi$ is compensated 
by higher order contributions in the chiral expansion \cite{brown962}, 
and therefore here the bare value is used, i.e. 
$\Sigma_{\mathrm{KN}}\rhos/\fps $.   
Compared to other chiral approaches \cite{weise93,li95} the 
resulting kaon dispersion relation shows a relatively strong 
density dependence. The increase of the 
in-medium $K^+$ mass ${\tilde m}_{\mathrm K}$, 
Eq. (\ref{effmass}), with this parameterisation 
is still consistent with the empirical 
knowledge of kaon-nucleus scattering and allows to explore 
in-medium effects on the production 
mechanism arising from zero temperature kaon potentials. 
For the kaon production via pion absorption $\pi B\longrightarrow YK^+ $ 
the elementary cross section of \cite{tuebingen1} are used. 
For the $N N \longrightarrow BYK^+ $ channels we apply 
the cross sections of Ref. \cite{sibirtsev95} which give a good fit 
to the COSY-data close to threshold. For the case of 
$N \Delta \longrightarrow BYK^+ $ and 
$\Delta \Delta \longrightarrow BYK^+ $ reactions experimental 
are rare. Thus we rely on the model calculation of ref. \cite{tsushima99}. 
In the case that 
a $N^*$ resonance is involved in the reaction we used the same 
cross section as for nucleons. 
In the presence of scalar and vector fields the kaon optical potential 
in nuclear matter has the same structure as the corresponding 
Schroedinger equivalent optical potential for nucleons \cite{dbhf}
\beq
U_{\rm opt}(\rho ,{\bf k}) 
=  -\Sigma_S + \frac{1}{\mk} k_{\mu} V^{\mu}  
+ \frac{\Sigma_S^2 - V_{\mu}^2}{2\mk}  ~.
\label{uopt}
\end{equation}
and leads to a shift of the thresholds conditions inside 
the medium. To fulfil 
energy-momentum conservation the optical potential is absorbed 
into an newly defined effective mass 
\beq
{\tilde m}_{\mathrm K} (\rho ,{\bf k}) 
= \sqrt{ \mks + 2\mk U_{\rm opt}(\rho ,{\bf k}) }
\label{effmass}
\eeq
which is a Lorentz scalar and sets the canonical momenta on the mass-shell 
$0=  k_{\mu}^{2} - {\tilde m}_{\mathrm K}^{2}$. 
Thus, e.g., the threshold condition for $K^+$ production in baryon induced 
reactions reads 
$\sqrt{s} \ge {\tilde m}_B + {\tilde m}_Y + {\tilde m}_K$ 
with $\sqrt{s}$ the centre--of--mass energy of the colliding baryons. 
For a consistent treatment of the thresholds the scalar 
and vector baryon mean fields entering into eq. 
(\ref{effmass}) are determined from two 
versions of the non-linear 
Walecka model with K=200/380 MeV, respectively \cite{li95b}. The hyperon 
field is thereby scaled by 2/3 which yields also a 
good description of the $\Lambda$ flow \cite{wang99b}. Since the 
parameterisations chosen for the non-linear Walecka model yield 
the same EOS as the Skyrme ones, the overall energy is conserved. 
The kaon production is treated perturbatively and does generally 
not affect the reaction dynamics \cite{fang94}.
\section{EOS dependence of $K^+$ production}
In Fig. 1 the $K^+$ excitation function for $Au+Au$ and $C+C$ 
reactions starting from 0.8 A$\cdot$GeV which is far below 
threshold ($E_{thr}=$ 1.58 GeV) are shown. The calculations are performed 
for a soft/hard EOS including the in-medium kaon potential. 
For both systems the agreement with the 
KaoS data \cite{kaos99,sturm00} is very good when a soft EOS is used. 
In the large system there 
is a visible EOS effect which is absent in the light system. 
To estimate the influence of the in-medium kaon potential for 
$C+C$ also calculations without potential are shown. 
Already in the light system the $K^+$ yield is reduced by about 
$50\%$ by the influence of the potential which is essential to 
reproduce the data \cite{kaos99}.  
\begin{figure}[h]
\unitlength1cm
\begin{picture}(8.,8.0)
\put(0,0){\makebox{\epsfig{file=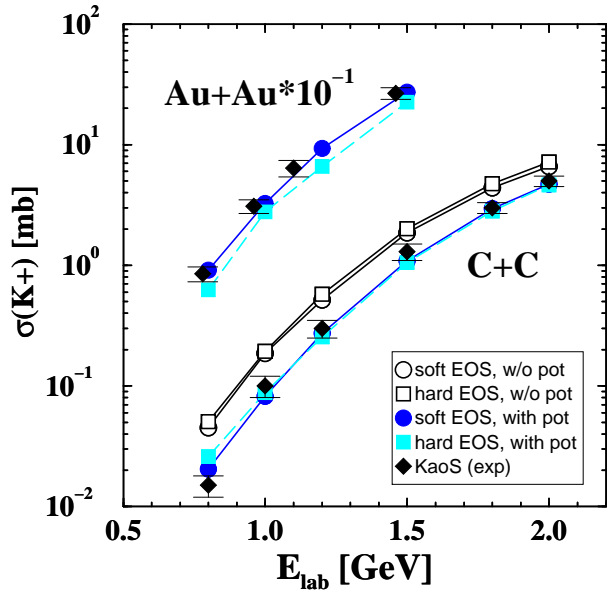,width=8.0cm}}}
\end{picture}
\caption{Excitation function of the $K^+$ production cross section 
in $Au+Au$ (scaled by $10^{-1}$) 
and $C+C$ reactions. The calculations are performed with 
in-medium kaon potential and using a hard/soft nuclear EOS 
and are compared to data from the KaoS Collaboration 
\protect\cite{kaos99,sturm00}. For $C+C$ also calculations without 
kaon potential are shown.  
}
\label{Fig1}
\end{figure}
To extract more clear information on the nuclear EOS 
next we consider the ratio $R$ of the kaon multiplicities 
obtained in $Au+Au$ over $C+C$ reactions, normalised to the corresponding 
mass numbers. In Fig. 2 central (b=0 fm) collisions are analysed. 
Going far below threshold $R$ strongly increases 
when a soft EOS is applied 
whereas the increase of $R$ is much less pronounced 
using the stiff EOS and $R$ even slightly drops at 
the lowest energy. 
\begin{figure}[h]
\unitlength1cm
\begin{picture}(8.,8.0)
\put(0,0){\makebox{\epsfig{file=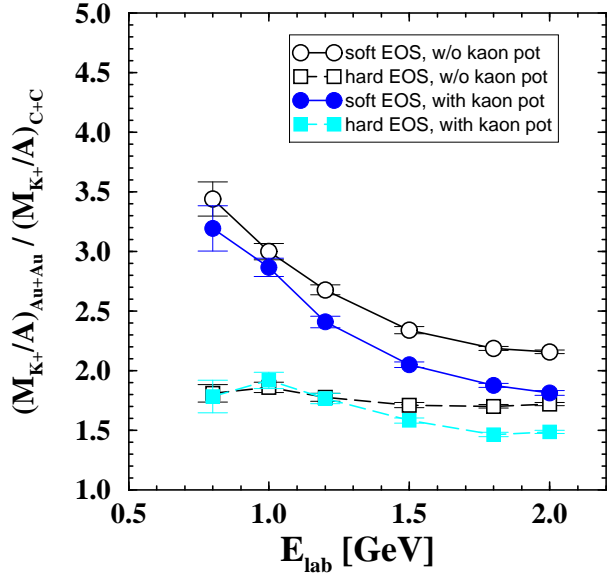,width=8.0cm}}}
\end{picture}
\caption{Excitation function of the ratio of $K^+$ 
multiplicities obtained in central  (b=0 fm) $Au+Au$ over $C+C$ 
reactions. The calculations are performed with/without 
in-medium kaon potential and using a hard/soft nuclear EOS. 
}
\label{Fig2}
\end{figure}
Hence, this ratio reflects the higher compression achieved 
in the heavy system. The $C+C$ system, on the other hand, 
is too small to develop a significantly larger compression in 
the case of a soft EOS compared to the hard EOS. 
Moreover, in the latter case the 
slightly more energetic binary collisions lead 
even to a higher $K^+$ yield in the case of a hard EOS 
at 0.8 A$\cdot$GeV. 
Remarkably, this behaviour is seen despite of the presence of 
an in-medium kaon potential which acts opposite to the EOS 
effect: a higher compression 
increases the kaon yield but also the value of the in-medium 
kaon mass which, on the other hand, tends to lower the yield again. 
However, the increase of the in-medium mass goes linear with 
density whereas the collision rate per volume increases approximately with 
$\rho^2$. E.g. in central $Au+Au$ reactions at 0.8 A$\cdot$GeV the 
average density $<\rho >$ at kaon production 
is enhanced from 1.47 to 1.57 $\rho_{\rm sat}$ switching from the 
hard to the soft EOS. This leads to an average shift of the in-medium 
mass (\ref{effmass}) compared to the vacuum value of 55/61 MeV using 
the hard/soft EOS, i.e. a relative shift of 6 MeV between 
soft and hard. However, collective effects 
like the accumulation of energy by multiple scattering 
show a higher sensitivity on the compression resulting in an 
enhancement of the available energy $<\sqrt{s}>=90$ MeV applying 
the soft EOS. For $C+C$ this effect is reverse ($<\sqrt{s}>=-45$ MeV) 
since the system is too small to develop a significant 
difference in compression 
and more repulsive collisions enhance the $K^+$ 
yield at low energies. This effect disappears above 1.0 A$\cdot$GeV.  
There exists thus a visuable EOS dependence of the kaon multiplicities. 
\begin{figure}[h]
\unitlength1cm
\begin{picture}(8.,8.2)
\put(0,0){\makebox{\epsfig{file=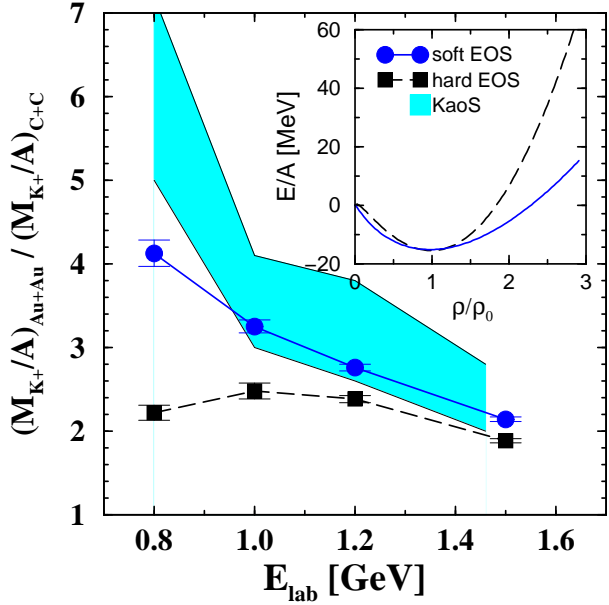,width=8.0cm}}}
\end{picture}
\caption{Excitation function of the ratio $R$ of $K^+$ 
multiplicities obtained in inclusive $Au+Au$ over $C+C$ 
reactions. The calculations are performed with 
in-medium kaon potential and using a hard/soft nuclear EOS and 
compared to the experimental range of $R$ 
(shaded area) given by the data from the 
KaoS Collaboration \protect\cite{sturm00}.
}
\label{Fig3}
\end{figure}
The comparison to the KaoS data \cite{sturm00} is finally made in Fig. 3. 
Here only calculations including the kaon potential are shown since 
it is already clear from Fig. 1 that without the potential 
one is not able to reproduce the experimental yields. The calculations 
are performed under minimal bias conditions with $b_{{\rm max}}=11$ fm 
for $Au+Au$ and $b_{{\rm max}}=5$ fm for $C+C$ and normalised to the 
experimental reaction cross sections \cite{kaos99,sturm00}. Both 
calculations show an increase of $R$ with decreasing incident energy 
down to 1.0 A$\cdot$GeV. As already seen for the central collisions 
this increase is much less pronounced using the stiff EOS. 
In the latter case $R$ even drops for 0.8 A$\cdot$GeV 
whereas the soft EOS leads to an unrelieved increase of $R$. 
At 1.5 A$\cdot$GeV which is already very close to threshold 
the differences between the two models tend to disappear. 
The overall behaviour of $R$ is found to be quite independent 
of the various production channels with initial states 
$ i= NN, \pi N, N\Delta, \pi\Delta, 
\Delta\Delta$. Ratios $R_i$ built separately 
for the individual channels show in both cases (soft or hard) 
a similar energy dependence as the total $R$ (except of $R_{\pi \Delta}$ 
which tends to remain large also at high energies). 
The transport calculations further 
demonstrate that the increase of $R$ is not  
is not caused by a trivial, i.e. EOS independent limitation 
of phase space at low energy in the small system. This is supported 
by the fact that the number of collisions which the involved particles 
encountered prior to the production of a kaon and which is a measure 
of the collectivity provided by the system does not reach a sharp 
limit for $C+C$ at low energies. 
The strong increase of $R$ can be directly related to 
higher compressible nuclear matter. The comparison to the experimental 
data from KaoS \cite{sturm00} where the increase of $R$ is even more 
pronounced strongly favours a soft equation of state. 
\section{Phase space for $K^+$ production}
To obtain a quantitative picture of the explored density 
effects in Fig. 4 the baryon densities are shown at 
which the kaons are created. The energy is chosen most below 
threshold, i.e. at 0.8 A$\cdot$GeV and only central collisions 
are considered where the effects are maximal. $dM_{K^+}/d\rho$ 
is defined as 
\beq
dM_{K^+}/d\rho = \sum_{i}^{N_{K^+}} 
\frac{d P_i }{ d\rho_B ({\bf x}_i, t_i)}
\eeq 
where $\rho_B$ is the baryon density at which the kaon $i$ was created 
and  $P_i$ is the corresponding production probability. For 
the comparison of the two systems the curves are normalised to the 
corresponding mass numbers. 
\begin{figure}[h]
\unitlength1cm
\begin{picture}(8.,8.2)
\put(0,0){\makebox{\epsfig{file=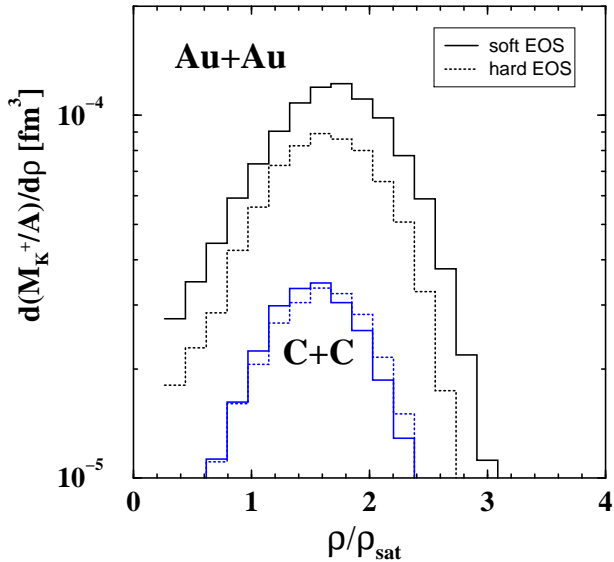,width=8.0cm}}}
\end{picture}
\caption{Kaon multiplicities (normalised to the mass numbers of the 
colliding nuclei) as a function of the baryon density 
at the space-time coordinates 
where the $K^+$ mesons have been created. Central (b=0 fm) $Au+Au$ and $C+C$ 
reactions at 0.8 A$\cdot$GeV are considered. 
The calculations are performed with 
in-medium kaon potential and using a hard/soft nuclear EOS. 
}
\label{Fig4}
\end{figure}
Fig.4 illustrates several features: 
Only in the case of a soft EOS the mean densities at which kaons 
are created differ significantly for the two different reaction 
systems, i.e. $<\rho /\rho_{\rm sat} >$=1.46/1.40 ($C+C$) 
and 1.47/1.57 ($Au+Au$) using  
the hard/soft EOS. Generally, in $C+C$ densities above 
$2\rho_{\rm sat}$ are rarely reached whereas in $Au+Au$ the kaons are 
created at densities up to three times saturation density. 
Furthermore, for $C+C$ the density distributions are weakly 
dependent on the nuclear EOS. The situation changes 
completely in $Au+Au$. Here the densities profile 
shows a pronounced EOS dependence \cite{li95b}. 
Moreover, the excess of kaons obtained with the soft EOS 
originates almost exclusively from 
high density matter which demonstrates that compression effects 
are probed. 
\begin{figure}[h]
\unitlength1cm
\begin{picture}(8.,9.5)
\put(0,0){\makebox{\epsfig{file=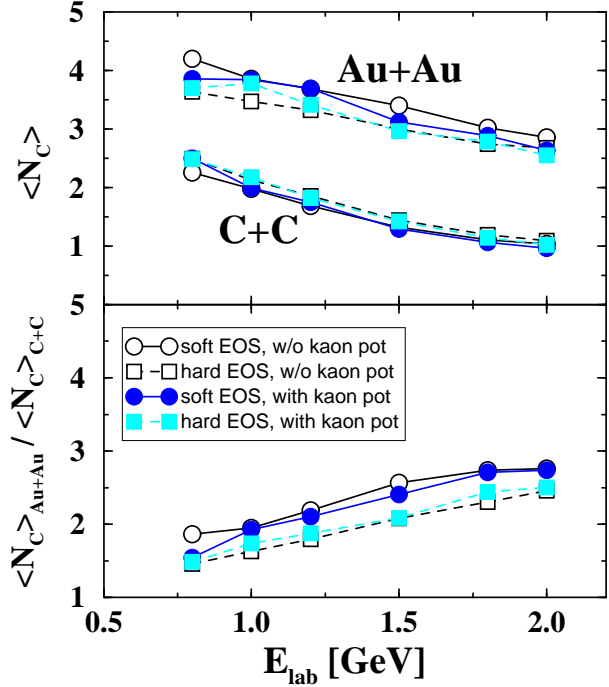,width=8.0cm}}}
\end{picture}
\caption{As a measure for the available phase space for $K^+$ 
production the mean number of collisions $<N_C >$ per particle 
which the hadrons ($N,\Delta,\pi$) did undergo before they produce 
a $K^+$ meson is considered. The upper panel shows $<N_C >$ in 
central $Au+Au$ and $C+C$ collisions. The lower panel shows the 
ratio of this quantity in $Au+Au$ over the same in $C+C$ 
reactions. The calculations are performed with/without 
in-medium kaon potential and using a hard/soft nuclear EOS.
}
\label{Fig5}
\end{figure}
Similar to the desnsity shown before a quantitative impression 
of the importance of collective and phase space 
effects can be obtained. Thus in Fig. 5 
the average number of collisions for those 
hadrons ($N, \Delta, \pi$) which were involved in the $K^+$ production 
are displayed. Again only central collisions are considered where the 
effects are maximal. $<N_C >$ is defined as 
\beq
<N_C > = \sum_{i}^{N_{K^+}} \frac{1}{2} (N_{C_{1}}^i + N_{C_{2}}^i) 
P_i / \sum_{i}^{N_{K^+}} P_i
\eeq 
with $ N_{C}^{i}$ being the number of collisions which particles  
($1,2$) experienced before they produced kaon $i$, and $P_i$ is the 
corresponding production probability. It is seen 
that in average the particles 
undergo about twice as much relevant collisions in the heavy 
compared to the light system. Furthermore, the collectivity, i.e. 
the accumulation of energy by multiple scattering, increases 
with decreasing incident energy. This feature is found to be 
independent of the system size. Thus one can conclude that the 
increase of $R$ is not due to a trivial phase space effect, namely 
the fact that far below threshold the $C+C$ system is  
simply too small to provide enough collectivity for 
kaon production. If such a scenario - which could model independently 
as well explain the rise of $R$ seen in the KaoS data - 
would be true, $<N_C >$ would have to saturate for $C+C$ at 
low energies. This is obviously not the case. 
Moreover, building also here the ratio (lower panel of Fig. 5) it seems that 
the relative enhancement of available phase space for $K^+$ 
production in the large system is decreasing at low energies. 
In the last figure we show $<N_C >$ (b=0 fm, SMD, with kaon pot.) 
for an even smaller system $Li^6$. Also here the number 
of NN-collisions involved in the production of a $K^+$ increases going 
far below threshold. This demonstrates that $K^+$ production 
far below threshold always requires a certain amount of collectivity 
which can be provided also in a very small colliding system, though 
such processes are extremely rare. There is, however, no sharp limit 
were such collision histories become impossible. 
Thus trivial phase space effects can be excluded 
for an explanation of the increase of $R$. 
\begin{figure}[h]
\unitlength1cm
\begin{picture}(8.,6.5)
\put(0,0){\makebox{\epsfig{file=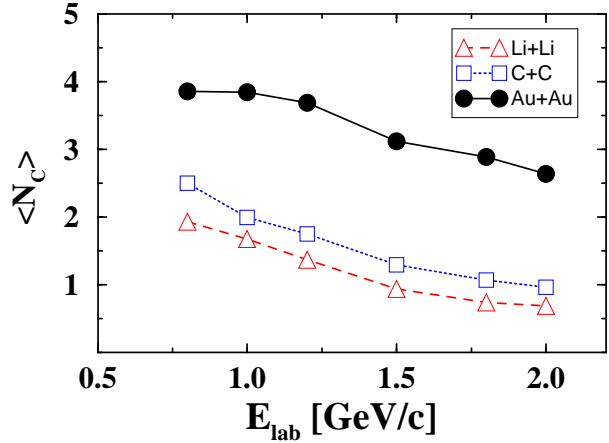,width=8.0cm}}}
\end{picture}
\caption{Mean number of collisions $<N_C >$ (see Eq.(4) ) 
for central  $Au+Au$, $C+C$ and $Li+Li$ collisions.
}
\label{Fig6}
\end{figure}
In \cite{sturm00} a similar argument was based on the measurement 
of high energy pions which can test the phase space available for 
paricle production. Astonishingly, the corresponding ratio buildt from 
high energy pions does not show any energy dependence but the 
excitation function remains almost constant. This feature 
contradicts a thermal interpretation where one expects particle 
abundances to be exclusively governed by phase space, independent from the 
fact if the necessary energy goes into mass or momentum. 
Above threshold $K^+$ production shows $m_t$ scaling according 
to available phase space, even if the particles are not produced 
from an equilibrated source \cite{brat98}. However, even if one 
assumes a thermal picture (which is, we believe that is does not 
hold for kaons at SIS energies since it is in contradiction to the 
knowledge from transport calculations) kaons and high $p_t$ pions 
should behave different below and around threshold. 
As shown in \cite{oeschler00} a thermal 
description at SIS energies must account for strangeness 
conservation in the $K^+$ production mechanism which, via the associated 
hyperons, introduces a dependence on the baryon chemical potential 
which is absent for the thermal pion yield. Thus $K^+$ production 
is coupled to the baryon density whereas high energy pions depend
only the temperature. This simple consideration 
illustrates that it can not a priori be excpected that high 
energy pions behave like kaons. However, the constant pion ratio 
supports our phase space arguments given above and the fact that 
non-trivial, i.e. EOS effects, are responsible for the strong 
increase of $R$ seen for the kaons. 
\section{Summary}
To summarise, we find that at incident energies far below 
the free threshold $K^+$ production is a suitable tool to study the 
dependence on the nuclear equation of state. Using a light system 
as reference frame there is a visible sensitivity on the EOS 
when ratios of heavy ($Au+Au$) over light ($C+C$) systems are considered. 
Transport calculations indicate that the $K^+$ production gets 
hardly affected by compressional effects in $C+C$ but is 
highly sensitive to the high density matter 
($1\le \rho /\rho_{\rm sat} \le 3$) created in $Au+Au$ reactions. 
Results for the  $K^+$ excitation function in $Au+Au$ over 
$C+C$ reactions as measured by the KaoS Collaboration, strongly 
support the scenario with a soft EOS. This statement is also 
valid when an enhancement of the in-medium kaon mass as predicted 
by chiral models is taken into account. 

The authors would like to acknowledge valuable discussions 
with J. Aichelin, H. Oeschler, P. Senger and C. Sturm. 



\end{document}